\documentclass[11pt]{article}
\def\fnum@figure{{\sc Fig.}\ \thefigure}
\long\def\@makecaption#1#2{
 \vskip 10pt 
 \setbox\@tempboxa\hbox{#1.\ #2}
 \ifdim \wd\@tempboxa >\hsize \unhbox\@tempboxa\par \else \hbox
to\hsize{\hfil\box\@tempboxa\hfil} 
 \fi}

\makeatletter
\def\section{\secdef\@sectionb\@sections}
\def\@sectionb[#1]#2{\@sections{#2}}
\def\@sections#1{\@startsection
  {section}{1}{\z@}{14pt}{14pt}{\centering\bf}{\uppercase{#1}}}
\def\subsection{\@startsection {subsection}{2}{\z@}{14pt}{14pt}{\bf}}
\def\subsubsection#1{\par\addvspace{14pt}{\bf #1.---\/}\ignorespaces}
\def\paragraph#1{{\it #1:\ \ \/}\ignorespaces}

\def\@citex[#1]#2{\if@filesw\immediate\write\@auxout{\string\citation{#2}}\fi
  \def\@citea{}\@cite{\@for\@citeb:=#2\do
    {\@citea\def\@citea{\@citesep}\@ifundefined
       {b@\@citeb}{{\bf ?}\@warning
       {Citation `\@citeb' on page \thepage \space undefined}}%
{\csname b@\@citeb\endcsname}}}{#1}}

\def\ps@top{\let\@mkboth\@gobbletwo
     \def\@oddhead{\rm\hfil\thepage\hfil}\def\@oddfoot{}
     \def\@evenhead{}\let\@evenfoot\@oddfoot}
\def\@bibsetup{\itemindent=-\leftmargin}
\def\@citesep{; }
\def\@cite#1#2{({#1\if@tempswa , #2\fi})}
\def\@biblabel#1{\hfill}
\def\thebibliography#1{\section*{References\markboth
 {REFERENCES}{REFERENCES}}\list
 {[\arabic{enumi}]}{\settowidth\labelwidth{[#1]}\leftmargin\labelwidth
 \advance\leftmargin\labelsep
 \usecounter{enumi}\@bibsetup}
 \def\newblock{\hskip .11em plus .33em minus -.07em}
 \sloppy
 \sfcode`\.=1000\relax}
\renewcommand{\section}{\@startsection {section}{1}{\z@}{-3.5ex plus -1ex minus 
    -.2ex}{2.3ex plus .2ex}{\centering\large\bf}}
\renewcommand{\subsection}{\@startsection{subsection}{2}{\z@}{-3.25ex plus
    -1ex minus -.2ex}{1.5ex plus .2ex}{\centering\bf}}
\makeatother
\newdimen\rotdimen
\def\vspec#1{\special{ps:#1}}
\def\rotstart#1{\vspec{gsave currentpoint currentpoint translate
   #1 neg exch neg exch translate}}
\def\rotfinish{\vspec{currentpoint grestore moveto}}
\def\rotr#1{\rotdimen=\ht#1\advance\rotdimen by\dp#1%
   \hbox to\rotdimen{\hskip\ht#1\vbox to\wd#1{\rotstart{90 rotate}%
   \box#1\vss}\hss}\rotfinish}
\def\rotl#1{\rotdimen=\ht#1\advance\rotdimen by\dp#1%
   \hbox to\rotdimen{\vbox to\wd#1{\vskip\wd#1\rotstart{270 rotate}%
   \box#1\vss}\hss}\rotfinish}%
\begin{document}
\vspace*{1in}
\begin{center}
{\large\bfseries N$_2^+$ and CO$^+$ in Comets 122P/1995\,S1 (deVico) and
C/1995\,O1 (Hale-Bopp)} \\ [25pt]
Anita L. Cochran, William D. Cochran and Edwin S. Barker \\
McDonald Observatory \\
University of Texas at Austin \\ [15pt]
Accepted for publication in {\itshape Icarus} \\ [2in]
\end{center}

\section{Abstract}
We observed comets 122P/1995\,S1 (deVico) and C/1995\,O1 (Hale-Bopp)
with high spectral resolving power in order to determine the
ratio of N$_2^+$/CO$^+$ in their comae.  While we clearly detected
the CO$^+$ in both of these comets, no N$_2^+$ was detected in either comet.
From these
spectra, we derive sensitive upper limits for N$_2^+$/CO$^+$.
These upper limits are substantially below other reported detections of
N$_2^+$/CO$^+$
in other comets.  We discuss the prior N$_2^+$ detections and compare
them with our observations.  The abundance of N$_2$ in comets is important
to our understanding of the condensation of ices in the solar
nebula.  In addition, N$_2$ is a tracer of Ar so study of N$_2$ allows
an understanding of the role of comets for delivering volatiles
to the terrestrial planets.  It appears that many, if not most, comets
are depleted in N$_2$ and it will be necessary to search for
a mechanism for depleting this molecule in order to be consistent
with current models of the solar nebula.

\newpage
\subsection{Introduction}

Nitrogen is one of the more abundant elements in the universe and is therefore
assumed to be an important constituent of the solar nebula and of
the comets.  Nitrogen probably exists in comets in the form of
N$_2$ and other nitrogen-bearing molecules, including NH$_{3}$.  
Indeed, the ratio N$_2$/NH$_{3}$ is a sensitive indicator of conditions
in the solar nebula.  Lewis and Prinn (1980) point out that
at high temperatures and low pressures the dominant equilibrium species
of carbon, oxygen and nitrogen would be N$_2$, CO, and H$_{2}$O.
Only as temperature and pressure regimes change will CH$_{4}$ and NH$_{3}$
be produced and N$_2$ and CO be depleted.  They conclude that
the conversion of N$_2$ to NH$_{3}$ and CO to CH$_{4}$ would be
sufficiently slow relative to radial mixing in the primitive solar
nebula so that only small amounts of NH$_{3}$ and CH$_{4}$ should be
present.  Conditions in the circumplanetary nebulae would be
sufficiently different so that jovian planets might have
increased NH$_{3}$ and CH$_{4}$ abundances (Prinn and Fegley 1981).
\nocite{lepr80,prfe81}

Comets delivered some of the volatiles that we see today in the atmospheres
of the terrestrial planets, but it is not certain how important a
source of volatiles the comets represent.  Owen and Bar-Nun (1995a,b)
have pointed out that N$_2$ is an important guide to the volatile 
abundances of comets because it is trapped and released by amorphous
ice in a manner which is similar to argon (Bar-Nun {\it et al.} 1988). 
\nocite{baklko88}
Using N$_2$ as a guide
to the argon, one can determine the extent to which comets enriched the
volatile and noble gas components of the terrestrial planets.
Ices formed at low temperatures
will trap gas from the surrounding nebula, fractionating the
original mixture as a function of the local temperature.  Thus, they
suggest that comets which formed near Uranus and Neptune, at temperatures
around 50\,K, would be the source of noble gases for Earth and Mars,
while the higher quantities of neon and argon in the atmosphere of Venus,
compared with Earth, would require
comets formed at colder temperatures, such as in the Kuiper belt, to
be the deliverers of some of the volatiles.
\nocite{owba95a,owba95b}

We detect such species as NH, NH$_{2}$ and CN in every comet, so evidence
of nitrogen carriers is easily available.   Most of these species and
their parents are chemically reactive in the comae of comets.
Molecular nitrogen should be less reactive than species
such as NH$_{3}$ or HCN.  
While spacecraft have flown past comet Halley with mass spectrometers onboard,
measurement of N$_2$ is difficult with mass spectrometry since both N$_2$
and CO occupy the mass 28 bin of these instruments (cf. Everhardt {\it et al.}
1987 for a discussion of CO and N$_2$ from Giotto observations of
Halley).  Thus, disentangling
the quantity of N$_2$ from the CO is very model dependent.
\nocite{ebetal87}

This leaves the field of ground-based spectroscopy for determining
the quantity of molecular nitrogen.
Ground-based studies of molecular nitrogen are very
difficult, however, because of the N$_2$ abundance of the Earth's
atmosphere.
To circumvent the difficulty in observing N$_2$, past observations have
concentrated on the N$_2$ ion, primarily through observations of
the N$_2^+$ (0,0) band at 3914\AA.  This band is extremely weak
and is expected to be seen only in the tails of comets.  Care must
be taken when observing this band since N$_2^+$ emission is also excited in
the atmosphere of the Earth, especially near dusk and dawn, when
comets are often observed.  Auroral activity will also excite this
band in the terrestrial atmosphere.  Additionally, this weak feature can easily
be confused with other, nearby, cometary emissions.  Thus, accurate
measurement of N$_2^+$ in cometary spectra requires both good spatial
and spectral resolution to separate the features from that of the Earth
and other cometary features.

\subsection{Observations}

We observed comets deVico and Hale-Bopp with the 2DCoude spectrograph (Tull
{\it et al.} 1995) on the 2.7-m Harlan J. Smith telescope of 
McDonald Observatory. \nocite{TuMQSn95}
The 2DCoude has two operating modes.  The ``lower" resolution mode
has a resolving power, R=60,000.  In this mode, spectral coverage
is complete from around 3800-5800\AA\ and coverage continues to 1\,$\mu$m
with increasing interorder gaps.  Typically, 60--65 spectral orders are
observed.  Therefore, in the blue, many molecular bands can be observed
simultaneously, regardless of the exact grating setting.
In the ``high" resolution mode, R=200,000, but
the coverage is much less complete than the lower resolution mode.
Typically, high resolution covers 10--15 orders of approximately 15\AA\ each.
Thus, care must be taken to center key features on a spectral order
and many features remain unobserved.

For this project, we observed comet Hale-Bopp in the high resolution
mode, carefully centering the portion of the order containing the N$_2^+$ band 
on the CCD, while also observing the CO$^+$ (2,0) and (3,0) bands on
two other orders.  Since the CH$^+$ (0,0) band occurs at a wavelength
coincident with the CO$^+$ (2,0) band, this ion was also observed.
Comet deVico was observed in the lower resolution mode and the same
three ions were observed.  Table~\ref{log} gives the circumstances of the
observations.  For all observations, the slit was 8.2\,arcsec long.
For the Hale-Bopp observations, the slit was 0.34\,arcsec wide, while
it was 1.2\,arcsec wide for deVico.  Each slit width projects to two
pixels on the CCD for the resolving power of the observations.
Different positions within the coma were observed by moving the telescope
around the sky under accurate computer control.

The data were reduced using the {\sc echelle} package of {\sc iraf}.
Incandescent lamp observations were used to determine the flat field;
ThAr lamp observations were used for calculating the dispersion curve.
The rms errors of our fits for the dispersion curve are 0.24\,m\AA\
for the Hale-Bopp spectra and 2.5\,m\AA\ for the lower resolution deVico
spectra.
The solar spectrum was observed with an identical instrumental setup
to that used for the comets by imaging the Sun through a diffuser
on the roof of the spectrograph slit room and projecting this image through
the slit in the same manner as objects observed through the telescope.
Thus, we used an {\itshape observed} solar spectrum in our reductions.

Care was taken to preserve the relative flux levels of the spectra.
The spectral orders were extracted by first tracing the order along
the chip and carefully setting the edges of the apertures.
Since the continua of the cometary spectra were rarely of high enough
signal/noise to define well the aperture boundaries, it was assumed
that the flat lamp boundaries were appropriate for the comet and only
the position on the chip of the center of the order was computed
for the cometary spectra.  Extraction was done using variance weighting.
We used a 1D fit for stars and solar spectra, while a 2D fit was used
for cometary spectra (because of the emission line nature of the cometary
spectra).  At the end of the routine reduction, we had files containing
$n$ spectra for each initial spectral image, where $n$ is the number
of extracted orders in the image (60 for deVico and 13 for Hale-Bopp).

The solar spectrum observations were used to remove
the underlying continuum from the cometary spectra.  Comet deVico
has very little solar continuum, but the continuum of Hale-Bopp was
quite strong.  We corrected the comet and the solar spectrum
for the geocentric and heliocentric Doppler shifts
so that both were on a common rest frame.  Then, the solar spectrum
was carefully weighted to match the continuum level of the comet
in regions away from cometary emissions.  Some scattered light might
still remain, but the amount is minimal and was removed when the
line intensities were calculated.

Figure~\ref{haleboppco+} shows the spectral order of the CO$^+$ (2,0)
and the CH$^+$ (0,0) bands in the spectrum obtained 100\,arcsec
tailward of the optocenter of Hale-Bopp.  For both these ions, the
predicted molecular transitions in the spectrum are marked.  Inspection of
this figure shows that only very low $J-$levels are observed
for CH$^+$, while slightly higher $J-$levels are observed
for CO$^+$.  However, even for CO$^+$, $J-$levels above 10 or 11
are not seen.  

Figure~\ref{devicoco+} shows the same spectral region for observations
100\,arcsec tailward of the optocenter of comet deVico.  
The spectral coverage of an order is longer at the lower resolving
power of the deVico observations.  Even with this larger coverage, only
one of the CO$^+$ (2,0) ladders is seen in these observations.
Both CH$^+$ and CO$^+$ are again present, though the ratio of CH$^+$/CO$^+$
may be slightly different in these two comets.

Figure~\ref{haleboppn2+} shows the Hale-Bopp spectrum obtained 10\,arcsec
tailward of the optocenter in the spectral order which should contain
the N$_2^+$ (0,0) band.  The N$_2^+$ transition is a
B\,$^2\Sigma$--X\,$^2\Sigma$
transition and therefore does not have a Q-branch.  The P- and R-branch
line positions marked are from Dick {\it et al.} (1978) and have an accuracy
of 0.01\,cm$^{-1}=2$\,m\AA.  Although the solar-subtracted spectrum is somewhat
noisy, there are no believable features.  There might be a spike
at 3909\AA\ and a broader spike at 3913.5\AA.  Neither of these
is coincident with any of the N$_2^+$ line positions. 
The errors in the wavelengths of our spectra, coupled with the N$_2^+$
laboratory errors would lead us to expect coincidence to 2\,m\AA.
\nocite{dietal78}
The most
believable feature is the broad feature starting at 3914.5\AA\ and degrading
redward.  The positions of the C$_{3}$ (0,2,0)-(0,0,0)
band transitions are marked underneath the spectrum. 
The feature seems to match well the
R-branch bandhead of this C$_{3}$ band.  Thus, we conclude
that if the feature is real, it is some residual C$_{3}$ emission.
Since this spectrum was obtained only 10,000\,km from the optocenter,
this does not seem an unlikely species to observe.  
Inspection of the spectrum obtained 100\,arcsec tailward of the optocenter
shows even less evidence for features.  We therefore conclude
that we did not detect {\itshape any} N$_2^+$ in the spectrum of
Hale-Bopp.

Figure~\ref{devicon2+} shows the comparable spectral region
for comet deVico, 100\,arcsec from the optocenter.  
There appears to an upward fluctuation between 3913.5 and 3915.1\AA, with spikes
at 3913.9 and 3914.9\AA.
However, at R=60,000, it is
impossible to tell if this feature is a molecular band and, if so, which
way it degrades, or to differentiate whether
it is C$_{3}$ or N$_2^+$. 
The N$_2^+$ P-branch bandhead occurs at 3914.3\AA\ and distinct lines of
the P-branch should be visible.  We do not detect lines at the
expected wavelengths, within the 3\,m\AA\ wavelength uncertainties.
On the strength of the Hale-Bopp
observation, this feature could be C$_{3}$.  With the
larger spectral coverage of the R=60,000 orders, the blue end of
this order contains the CN (0,0) bandhead (not shown in Fig.~\ref{devicon2+}). 
Thus, we were able to
use the high signal/noise CN emission lines to confirm that there were no
errors in our wavelength solution or in our Doppler shift corrections.
The centers of the CN lines fell at the correct wavelengths,
verifying that any spikes in the N$_2^+$ region would also have
to be at predicted wavelengths.

For the deVico observations, our only spectrum off the optocenter
was obtained more than 70,000\,km into the tail.  Thus, it is reasonable to
ask about the likelihood that we observed C$_{3}$ this far from the optocenter.
Figure~\ref{devicoopt} shows the optocenter spectrum from the
same night.  We show more of the order to illustrate the abundance
of molecular emissions observed.  The CH B--X (0,0) band is clearly detected
along with several C$_{3}$ bands.  Indeed, inspection of this plot
shows there are several broad, unidentified features whose structure
seems similar to the identified C$_{3}$ bands.  Thus, it is likely
that this order is riddled with C$_{3}$. 
However, comparison of the strength of the strongest C$_{3}$ band, the
(0,0,0)~--~(0,0,0) band, in the optocenter and tail spectra makes it 
unlikely that we detected the much weaker (0,2,0)~--~(0,0,0) R-branch
bandhead in the tail spectrum.  Thus, it is most likely we detected only
noise in the deVico spectrum.

In summary, we do not believe that any N$_2^+$ was detected in the
spectra of comets Hale-Bopp or deVico.
The CO$^+$ and CH$^+$ were clearly detected
in both comet's spectra.  We are therefore able to place limits
on the important ratio of N$_2^+$/CO$^+$ in these two comets.

\subsection{Limits on N$_2^+$/CO$^+$}

For both comets Hale-Bopp and deVico, the CO$^+$ (2,0) band was clearly
detected.  Thus, we can derive an abundance of CO$^+$ from these
data for comparison with other comets.  However, coud\'{e} data can
not be easily calibrated into absolute fluxes, so we must work with band
intensities in detector counts. 
In addition, typical lower resolving power observations
observe all branches of the CO$^+$ band, while we only observe
the $^2\Pi_{1/2}$(F$_2$) branches.  We took the simple approach
of ``integrating'' the band by fitting a continuum and summing
the counts in the band above the continuum.  We limited
our bandpass to just the region of the detected lines.
These values are given in Table~\ref{results}.
While a larger bandpass would be more comparable to prior low-resolution
observations of comets, it is  inappropriate for high resolution
observations since larger bandpasses would increase the noise with
no increase in signal.  Typical low-resolution observations do not
return to continuum in between the lines of different bands.

We do not believe that we have detected the N$_2^+$ in either
comet.  We computed upper limits by computing how
much of a band could be hidden within the noise.  
We did this by computing the rms in a bandpass. Then, the
upper limit is just $1/2 \times \mathrm{rms} \times \mathrm{bandpass}$,
in appropriate units.
These are $2\sigma$ upper limits.
For the same rms, more signal can be hidden in a large bandpass than
a small bandpass.  Since we do not know exactly how many lines would be
likely to be detected, we cannot easily define the bandpass.
We assumed a bandpass which would include the complete P-branch of N$_2^+$.
The derived counts, which
are $2\sigma$ upper limits for what could be hidden in the noise, are
listed in Table~\ref{results}, column 3.

With the use of a few assumptions and simplifications, we can
use our values to derive N$_2^+$/CO$^+$ for these two comets.
Examination of the solar spectra in these two regions, in comparison
with the published atlas of Kurucz {\it et al.} (1984), \nocite{kufubr84}
shows that the sensitivity of the N$_2^+$ order is lower than that of
the CO$^+$ order.  To match their sensitivity, we would need to
multiply the N$_2^+$ upper limits by a factor of 1.7.
However, the calibration of the solar spectrum depends
on details such as activity, so this factor is uncertain.  We do have
observations of $\alpha$ Lyr with the same instrumental setup
as for deVico, but since the N$_2^+$ band occurs in the Balmer decrement,
where the $\alpha$ Lyr flux changes rapidly with wavelength,
the $\alpha$ Lyr flux is not calibrated in this region.
Assuming a smooth decrease through the Balmer decrement, we confirm that a
correction factor of 1.5--1.7 for the N$_2^+$ counts would be appropriate.
We therefore adopt a factor of 1.6.

Once the band intensity is known, the column density can be
computed using
\[ N = L/g_{\nu^\prime\nu^{\prime\prime}} \]
where N is the column density, L is the integrated band intensity and
$g_{\nu^\prime\nu^{\prime\prime}}$
is the excitation factor.  We used excitation factors of
$7.0\times10^{-2}$\,photons\,sec$^{-1}$\,mol$^{-1}$ for the N$_2^+$ (0,0) band
(Lutz {\it et al.} 1993) and $3.55\times10^{-3}$\,photons\,sec$^{-1}$\,mol$^{-1}$
for the CO$^{+}$ (2,0) band (the average value from Figure~2 of
Magnani and A'Hearn 1986).
\nocite{maah86,luwowa93}
Then,
\[ \frac{\mathrm{N}_2^+}{\mathrm{CO}^+} =
         \frac{g_{\mathrm{CO}^+}}{g_{\mathrm{N}_2^+}}
         \frac{\mathrm{L}_{\mathrm{N}_2^+}}{\mathrm{L}_{\mathrm{CO}^+}} \]

For CO$^+$, we observed only one of the two ladders.  If we assume
the two ladders are equal strength, we should multiply our CO$^+$
intensity by two for the calculation.  We likewise need
to multiply the N$_2^+$ upper limits by a factor of two since we
have only measured the P-branch and the R-branch should have a similar
intensity.  In Table~\ref{results}, column 4, we list our upper limits
for N$_2^+$/CO$^+$, including using a factor of 1.6 to correct
for the sensitivity difference of the two orders.

It would be impossible to hide much N$_2^+$ in our spectra.
Figure~\ref{fake} (upper panel) shows one of the Hale-Bopp spectra,
as observed, and, in the lower panel, the same spectrum with a feature
added which has enough
integrated counts to yield the Halley N$_2^+$/CO$^+$ ratio
(Wyckoff and Theobald 1989 -- discussed below).
We do not claim that this synthetic band is the exact shape that would be
present, nor are the ``lines'' at exactly the N$_2^+$ wavelengths,
but it gives an idea of the ease with which we would detect such a feature.
Clearly, no feature this distinctive could be missed in our observations.
\nocite{wyth89}

\subsection{Previous Observations of N$_2^+$}

Most comets are not bright enough to be observed with the high spectral
resolving powers that we used for deVico and Hale-Bopp.  This was
especially true in the past, when detectors, such as photographic
plates, had much lower quantum efficiency than our current CCD detectors.
Therefore, prior observations which have detected N$_2^+$ in cometary
spectra have been obtained with lower resolution, often on photographic
plates.

The N$_2^+$ feature is generally weak and is overwhelmed by other molecular
emissions near the optocenter.  In addition, since N$_2^+$ is an ion,
it is entrained in the solar wind magnetic field and rapidly accelerated
into the tail.
Thus, spectra of the tail region are necessary for its definitive
detection, yet tail spectra are generally of lower signal/noise
than near-optocenter spectra since the cometary brightness falls with
increasing cometocentric distance.  Despite these difficulties,
observations of comets exist which show the
detection of N$_2^+$ in the tails of comets.

Only two of the prior reported observations are digitally measured spectra; the
rest are estimates from digital spectra or are photographic spectra.
Wyckoff and Theobald (1989) report a detection of N$_2^+$ in the
tail of comet Halley at a cometocentric distance of $3\times10^5$\,km
tailward.  These observations were at much lower resolution
than our observations.  They detected a weak emission in the region from
3885--3950\AA\ which they concluded was composed of contributions
from the CO$^+$ (5,1), CO$_2^+$ (unassigned), N$_2^+$ (0,0) bands and an
unidentified band. By modeling
the combined feature, they were able to estimate the contribution
of N$_2^+$ to the mixture.
Using this estimate and the average for the CO$^+$ (2,0), (3,0) and (4,0)
column densities, they derived a value of N$_2^+$/CO$^+ = 0.004$.
However, the excitation factor which Wyckoff and Theobald used
for N$_2^+$ was not accurate and Wyckoff {\it et al.} (1991b)
revised the value of the column density of N$_2^+$ using the
excitation factors of Lutz (1989--a personal communication).  This excitation
factor is the same as that
given in Lutz {\it et al.} (1993).  If we apply the value
from Lutz {\it et al.}, then N$_2^+$/CO$^+ = 0.002$.  If
only the (2,0) band column density of CO$^+$ is used, then
N$_2^+$/CO$^+ = 0.003$.
\nocite{wyteen91b,wyth89,luwowa93}

Lutz {\it et al.} (1993) reported observations of the tails of two
comets obtained at low resolution ($\sim10$\AA).  For comet Halley,
they obtained spectra at $2\times10^4$ and $2\times10^5$\,km from
the optocenter in the tailward direction.  They claim to have detected
no N$_2^+$ emissions in the Halley tail spectra.  However, they
also did not detect the CO$^+$ emissions in several Halley spectra.
Their derived upper limits for N$_2^+$/CO$^+$ when CO$^+$ was detected
were higher than the Wyckoff and Theobald detection.

In addition to observations of Halley, Lutz {\it et al.} also observed
comet C/1987\,P1 (Bradfield=1987 XXIX).  For this comet, spectra were
obtained at $2\times10^4$ and $6\times10^4$\,km from the optocenter.
CO$^+$ was detected in both spectra, but N$_2^+$ was only detected
at the larger cometocentric distance.  At their resolution, the
N$_2^+$ feature is on the wing of the CN (0,0) band.  No mention is
made of the possible contamination of this feature by the CO$^+$ (5,1)
band.   Assuming all of their measured band was N$_2^+$,
Lutz {\it et al.} derive a value of N$_2^+$/CO$^+ = 0.02$.

The vast majority of observations of cometary tails were photographic.
Not only were they at lower resolving powers than our observations
of deVico and Hale-Bopp, but photographic plates are even more difficult
to calibrate!  Non-uniformity in response and vignetting of the spectrograph
slit cause difficulty interpreting these spectra.  Still, there are
many fine examples of photographic spectra and these can be used
to determine the N$_2^+$/CO$^+$ ratio for some comets.
The largest published collection of photographic spectra is that
of Swings and Haser (1956).  
Examination of the plates in this atlas shows some comets for which
CO$^+$ and N$_2^+$ are both apparent, while other comets show
evidence of tails (i.e. CO$^+$) but no N$_2^+$.  
Arpigny examined these and other photographic and digital spectra at his
disposal and
estimated the intensity ratio for the N$_2^+$/CO$^+$, where
the CO$^+$ band used was the (4,0) band because of its proximity
to the N$_2^+$ emission (1999, personal communication).
Table~\ref{claude} lists the 12 comets which he determined had
both N$_2^+$ and CO$^+$ in these spectra, along with his
estimate of the ratio of the intensity of the two bands (column 2).
In order to compare his intensity ratios with other observer's
column density ratios, it is necessary to multiply by the
ratio of the excitation factors, as before.  Since the (2,0) CO$^+$
was used in our work and in other published ratios, we
converted the intensity ratios in Table~\ref{claude} by using the
relationship $I(4,0) = 0.6 \times I(2,0)$, where $I(4,0)$ is the
intensity of the (4,0) band, $I(2,0)$ is the intensity of the (2,0) band,
and the factor is taken from Table~4 of Magnani and A'Hearn (1986).
The resultant column density ratios are given in column~3 of
Table~\ref{claude}. 
Arpigny's estimates of the intensity ratios are consistent
with the published numbers for comet Bester (Swings and Page 1950)
and comet Humason (Greenstein 1962).  It should be noted
that Warner and Harding (1963) also observed comet Humason at a comparable
heliocentric distance (however they only discuss CO$^+$, not N$_2^+$).
\nocite{swpa50,gr62,waha63}\nocite{swingsatlas}
However, it is clear from Arpigny's compilation that N$_2^+$ has been observed
in previous spectra of some comets.

In addition, Arpigny reported four comets which had good spectra
but for which no, or only very faint, evidence of a plasma tail
existed.  These comets are C/1948\,V1 (Eclipse), C/1963\,A1 (Ikeya),
C/1968\,N1 (Honda), and C/1975\,N1 (Kobayashi-Berger-Milon).
Arpigny points out that N$_2^+$ emission is always very weak, so we should
not expect to see it when the CO$^+$ is weak or non-existent.

Our own examination of the atlas of Swings and Haser (1956) found
five comets for which there was evidence of a tail but no evidence for
N$_2^+$.  The Big Comet of 1910 (1910\,I) showed only continuum
in the tail, so this was presumably a dust tail.  Comets Halley (1910\,II),
Brooks (1911\,V), Gale (1912\,II), and Jurlof-Achmarof-Hassel (1939\,III)
showed evidence of weak CO$^+$ emission but no N$_2^+$ emission 
(Arpigny notes that N$_2^+$ was observed by Bobrovnikoff in spectra
of Halley obtained in 1910, but these spectra are not included
in the Swings and Haser Atlas).  These would
be similar to Hale-Bopp and deVico in the absence of N$_2^+$ while
other ions are present.
However, with the weakness of the CO$^+$ emissions in these four
photographically observed comets, the N$_2^+$ emission is most probably
below the plate sensitivity.

\subsection{Implications}

In this paper, we have presented high resolution observations of two
comets with which we were able to study the relative abundances of N$_2^+$ and
CO$^+$.  These two ions are proxies for understanding the quantity
of N$_2$ and CO, two of the least chemically reactive cometary coma species.
Conversion from the quantity of the ions to the quantity of the neutrals
is dependent on an understanding of the photodestruction branching ratios 
which are not well understood (Wyckoff and Theobald argue you must multiply
the ion ratio by a factor of 2, while Lutz {\it et al.} find no factor
necessary), so we will continue to discuss these
species in terms of their ions.  For both Hale-Bopp and deVico, CO$^+$
was easily detected but N$_2^+$ appears to be missing from the spectra.
Thus, we have put very low upper limits on the ratio of N$_2^+$/CO$^+$.
We note, however, that there
are previous observations of comets which show CO$^+$ but not N$_2^+$ for
which sensitive upper limits cannot be derived.

The quantity of N$_2$ and CO expected in a comet depends on several
factors including the temperature at which the ice was deposited, when
in the history of the formation of the solar system the gases were
trapped in the ice and the orbital history of the comet itself.
Current models of the solar nebula have comets which now
reside in the Oort cloud forming in the Uranus-Neptune region
(cf. Weissman 1991; Duncan, Quinn and Tremaine 1987).
\nocite{we91,duqutr87} 
The temperature in this region was probably about $50\pm20$\,K (Boss
{\it et al.} 1989).  \nocite{bomots89}
Thus, a first guess to the deposition temperature of cometary ices is 50K.
This is consistent with laboratory experiments described by Owen
and Bar-Nun (1995b).  \nocite{owba95b}

The first direct measure of a deposition temperature for ice came
with observations of deuterium in comet Hale-Bopp.
Meier {\it et al.} (1998b) reported the detection of HDO in Hale-Bopp
and determined a ratio of D/H=$(3.3\pm0.8)\times10^{-4}$ in H$_{2}$O.
In addition, Meier {\it et al.} (1998a) detected DCN for the first time
and derived a ratio of D/H=$(2.3\pm0.4)\times10^{-3}$ in HCN.
Note that the D/H ratio is different for these two species, with
D/H measured from HCN 7 times higher than from H$_{2}$O.
Since the D/H enrichment for different molecules is a strong function
of temperature, Meier {\it et al.} (1998a) were able to derive
a temperature for the cloud fragment in which this comet formed of
{\itshape no colder than} $30\pm10$\,K.
\nocite{meetal98a,meetal98b}

Bar-Nun {\it et al.} (1988) performed laboratory experiments on deposition
of various gases along with H$_{2}$O ice and
showed that CO is trapped 20 times more efficiently than N$_2$ in amorphous ice
which formed at 50\,K, when these two gases are present in equal
abundances with CH$_{4}$ and Ar.  This ratio changes slightly when
only CO and N$_2$ are present in the gas (Notesco and Bar-Nun 1996, Table~I)
but generally shows enrichment factors of 15--30.  From these laboratory
experiments, Owen and Bar-Nun (1995a) concluded that icy planetesimals
formed in the solar nebula at around 50\,K, the temperature at which
the studied comets should have formed, would have N$_2$/CO$\approx 0.06$
in the gases trapped in the ice if N$_2$/CO$\approx 1$ in the nebula.
The predicted cometary ratio of N$_2$/CO is much higher than our upper
limits for deVico
and Hale-Bopp and is higher even than the detections of Wyckoff and
Theobald (1989) and Lutz {\it et al.} (1993), though some of the
estimates might show ratios this high.
\nocite{baklko88,noba96,owba95a,wyth89,luwowa93}

Several factors might mitigate this discrepancy.  Prialnik and Bar-Nun (1990)
point out that the gas/water vapor ratio is not necessarily representative
of the ratio of ices in the nucleus.  However, the laboratory experiments
of Bar-Nun {\it et al.} (1988) have demonstrated that CO and N$_2$ should
be released simultaneously in the same proportion as they exist in the
ices.  There is evidence for a source of CO
at around 10,000\,km from the nucleus (Eberhardt {\it et al.} 1987) which
may be attributable to grains.  In addition, Krankowsky (1991) points out that
H$_2$CO is probably an additional parent for CO.  Thus, there may be
additional mechanisms for the production of CO which do not exist
for N$_2$.
\nocite{ebetal87,kr91}
While these factors might change the predicted ratio for N$_2$/CO,
Owen and Bar-Nun (1995a) go on to make the specific prediction
that ``future observations of dynamically new comets will show
values of N$_2$/CO systematically higher than those in well-established
short-period comets''.  They point out that as comets are continuously
exposed to solar radiation in the inner solar system, they would be
expected to lose any N$_2$ in their outer layers in a brief period of time.

Figure~\ref{values} shows all of the values and limits discussed in this
paper, plotted as a function of 1/a$_o$, the original semimajor axis
[four of the
values for the estimates are the oscullating 1/a,
as noted in Table~\ref{claude}; for C/1987\,P1 (Bradfield)
1/a$_o$= 0.006380 (Marsden and 
Williams 1995); for Hale-Bopp 1/a$_o$=0.00535 and for deVico 1/a$_{osc}$=0.057
(Marsden personal communications 1999)]. 
\nocite{mawi95}
For the Wyckoff and Theobald Halley observation, we include
the range of derived values.

If one ignores the Hale-Bopp upper limit, there would seem to be
an increase of N$_2^+$/CO$^+$ with decreasing 1/a$_o$, as was
predicted by Owen and Bar-Nun.  However, the trend is based mostly
on Arpigny's estimates, which are approximate.
In addition, the Hale-Bopp upper limit can not be discarded since
it is clear from inspection of the spectrum in Figure~\ref{fake}
that even as much N$_2^+$ as would be needed to equal the Halley
N$_2^+$/CO$^+$ cannot be hidden in this spectrum.
Conversely, while the \underline{values} for the ratio can be questioned for
the estimates, inspection of the atlas of Swings and Haser (1956)
and spectra such as the Humason spectrum of Greenstein (1962) 
clearly show an emission which is coincident with the location of 
the N$_2^+$ feature.  Thus, at least some comets have 
some N$_2^+$ in their spectra.
However, at lower spectral resolution, blending of features may lead to spurious
detections of N$_2^+$ and wrong estimates of the strength of this band.
The potential for blending, coupled with the weakness of the
N$_2^+$ feature even when it exists, point to a need for caution in
interpretation.

Thus, at this point, we have contradictory evidence for the ratio of
N$_2$/CO.  At least in the active outer regions of Hale-Bopp and
deVico, these comets appear to be very depleted in N$_2$ relative to CO.
The observations of Halley
by Wyckoff and Theobald (1989) also pointed to a depletion of N$_2$ for
Halley.  Indeed, Wyckoff {\it et al.} (1991a) derived a nitrogen
depletion for comet Halley of a factor of $\sim6$ relative to the
Sun.  \nocite{wyteen91}
Owen and Bar-Nun (1995) have pointed out the strong temperature dependence
of trapping of N$_2$ and CO.  Comets for which the deposition temperature
was greater than 50K could trap progressively less CO and N$_2$.
However, H$_2$CO will continue to be trapped in comparable quantities to
ices deposited at 50K.  Thus, there would continue to be a source
of CO, but the N$_2$ will be depleted relative to the CO.

Perhaps the solution to the quandary of depleted nitrogen is that our
assumption that the solar nebula
preferentially condenses nitrogen into N$_2$ instead of NH$_{3}$ is incorrect.
However, observations of dense molecular clouds (Womack {\it et al.} 1992)
have shown that N$_2 >>$~NH$_{3}$ for these potential star-forming sites.
\nocite{wowyzi92}
Another possibility is that the comets formed with much more molecular nitrogen
but that it was depleted post-formation.  It is certainly true
that comets will deplete volatile gases in their outer layers
as they pass close to the Sun, but this cannot be used as an explanation
when comparing comets with similar orbital histories, such as
Halley and deVico or Hale-Bopp and Bennett and C/1987P1 (Bradfield),
which show discrepant ratios of N$_2^+$/CO$^+$. 
Indeed, Engel {\it et al.} (1990) conclude that there might be some
post-formational processing of Halley, but not to any large extent
because of the low internal temperatures which are derived from the spin
temperature of H$_{2}$O (Mumma {\it et al.} 1993). \nocite{muwest93}
\nocite{enlule90}
They point out, however, that gas can be trapped in water ice efficiently,
but only if the ice is amorphous, such as it is in the various laboratory
experiments.  They conclude that codeposition into amorphous ice is
unlikely to be a favorable mechanism in forming cometary ices
since the water ice condenses, for most solar nebula models,
at 140--160\,K, at temperatures where the ice is likely to be
crystalline and to not adsorp volatiles readily.
However, the D/H ratios of H$_{2}$O and HCN suggest that cometary
deposition temperatures were not as warm as Engel {\it et al.} posit.

In summary, in this paper we presented evidence of two comets for which
no N$_2^+$ was detected, along with stringent upper limits, which would
indicate that these comets are depleted in N$_2$ relative to CO.
These observations are at odds with our understanding of the formation
processes of ices in the solar nebula.  Either a mechanism must be
found to deplete the N$_2$ ice once formed or we must understand
how a gaseous cloud with N$_2 >>$~NH$_{3}$ formed ices which
do not contain much molecular nitrogen.  Since we believe that
N$_2$ will be deposited into H$_{2}$O ice in a manner which is similar
to Ar, understanding this process has important implications
for understanding the role of comets for delivery of noble gases to
the terrestrial planets.  It is therefore important that more unambiguous
observations of the ion tails of comets be obtained, when possible,
to determine the intrinsic values of N$_2$/CO in comets.

\vspace{1in}
\begin{center}Acknowledgements\end{center}
This work was funded by NASA Grant NAG5 4208.
We thank Walter Huebner for encouraging the Hale-Bopp observations and
Toby Owen for stimulating our examination of the data.
We especially thank Claude Arpigny for graciously allowing us to use
his estimates of earlier observations and for many helpful discussions.


\clearpage
\begin{table}
\vspace{1in}
\setbox0=\vbox{
\begin{minipage}{6.5in}
\begin{center}
\caption{Observational Parameters\label{log}}
\begin{tabular}{lccccccccc}           
 \\ [-5pt]
\hline
 & & & & & & \multicolumn{2}{c}{Distance} \\
\multicolumn{1}{c}{Comet} & Date & R$_h$ & $\Delta$ & \.{R}$_h$ & $\dot\Delta$ &
\multicolumn{2}{c}{Tailward} & Start & Exposure \\
\cline{7-8}
 & & ({\sc au}) & ({\sc au}) & (km/sec) & (km/sec) & (arcsec) & (km) & (UT) & (sec) \\
\hline
 \\ [-8pt]
deVico & 3 Oct 1995 & 0.66 & 1.00 & -3.4 & -14.3 & 0 & optocenter & 11:37 & 600 \\
 & & & & & & 0 & optocenter & 11:57 & 600 \\
 & 4 Oct 1995 & 0.66 & 0.99 & -2.3 & -12.9 & 0 & optocenter & 11:11 & 1500 \\
 & & & & & & 100 & 71,775 & 11:42 & 1200 \\
 \\ [-5pt]
Hale-Bopp & 6 Apr 1997 & 0.92 & 1.40 & 2.9 & 18.4 & 0 & optocenter & 1:37 & 120 \\
 & & & & & & 10 & 10,150 & 1:46 & 900 \\
 & & & & & & 100 & 101,500 & 2:25 & 1800 \\
\hline
\end{tabular}
\end{center}
\end{minipage}}\rotl{0}
\end{table}

\clearpage
\begin{table}
\begin{center}
\caption{Results}\label{results}
\begin{tabular}{l|c|c|r}   
\multicolumn{4}{c}{ } \\ [-5pt]
\hline
\multicolumn{1}{c|}{Comet} & \multicolumn{1}{c|}{CO$^+$} & \multicolumn{1}{c|}{N$_2^+$} & \multicolumn{1}{c}{{N}$_2^+$/CO$^+$} \\
 & \multicolumn{1}{c|}{(counts)} & \multicolumn{1}{c|}{Upper Limit$^a$} &
\multicolumn{1}{c}{Upper Limit$^b$} \\ 
\hline
  & & & \\ [-8pt]
deVico & 718 & 2.7 & $3.0\times10^{-4}$ \\
Hale-Bopp$^c$  & 1434 & 1.7 & $9.9\times10^{-5}$ \\
Hale-Bopp$^d$  & 1306 & 1.0& $6.5\times10^{-5}$ \\
\hline
\multicolumn{4}{l}{\underline{Notes:}} \\
\multicolumn{4}{l}{\ \ $^a$ 3910.9--3914.33\AA\ bandpass} \\
\multicolumn{4}{l}{\ \ $^b$ Corrected for order sensitivity (see text)} \\
\multicolumn{4}{l}{\ \ $^c$ 10 arcsec tailward} \\
\multicolumn{4}{l}{\ \ $^d$ 100 arcsec tailward} \\
\end{tabular}
\end{center}
\end{table}

\clearpage
\hspace*{-10pt}
\begin{table}
\centering
\caption{Estimates from Earlier Spectra}\label{claude}
\begin{tabular}{llccrl}
\\
\hline
 && Intensity & Molecules & \multicolumn{1}{c}{1/a$_o$} \\
\multicolumn{2}{c}{Comet$^a$} &
N$_2^+$/CO$^+$ (4,0) $^b$ & N$_2^+$/CO$^+$ & \multicolumn{1}{c}{({\sc au}$^{-1})$ $^c$} &\multicolumn{1}{c}{Type$^d$}  \\
\hline
C/1908\,R1 & Morehouse (1908\,III)             & $\geq0.7$ & $\geq0.06$ & 0.000174 & pg\\
C/1940\,R2 & Cunningham (1941\,I)              & $\geq0.5$ & $\geq0.04$ & 0.000001 & pg\\
C/1947\,S1 & Bester (1948\,I)                  & 0.6--1.0  & 0.05--0.09 & 0.000024 & pg\\
C/1956\,R1 & Arend-Roland (1957\,III)          & $>1$      & $>0.09$    & -0.000531 & pg \\
C/1957\,P1 & Mrkos ( 1957\,V)                  & 0.2       & 0.02       & 0.002001 & pg\\
C/1961\,R1 & Humason (1962\,VIII)              & 0.2--0.3  & 0.02--0.03 & 0.004935 & pg\\
C/1969\,T1 & Tago-Sato-Kosaka (1969\,IX)       & $\leq0.3$ & $\leq0.03$ & 0.000507 & pg\\
C/1969\,Y1 & Bennett (1970\,II)                & 0.1--0.3  & 0.008--0.03& 0.007082 & pg\\
C/1973\,E1 & Kohoutek (1973\,XII)              & 0.8       & 0.07       & 0.000020 & FTS\\
C/1975\,V1-A & West (1976\,VI)                   & 0.1       & 0.008    & 0.001569 & IT\\
1P/1982\,U1 & Halley (1986\,III)                & $<0.1$    & $<0.008$   & 0.055737 & pg,\\
            &                                  &            &           &           & CCD, ret \\
C/1986\,P1 & Wilson (1987\,VII)                & 0.8       & 0.07       & -0.000260 & CCD\\
\hline
Notes: \\
\multicolumn{6}{l}{\hspace*{2em}$^a$ Old designations listed in parentheses besides name} \\
\multicolumn{6}{l}{\hspace*{2em}$^b$ Claude Arpigny -- personal communication} \\
\multicolumn{6}{l}{\hspace*{2em}$^c$ From Marsden and Williams (1995)} \\
\multicolumn{6}{l}{\hspace*{2.5em} 1/a (osculating) for Arend-Roland, Bennett, Halley and Wilson} \\
\multicolumn{6}{l}{\hspace*{2em}$^d$ pg=photgraphic;FTS=Fourier Transform
spectrometer; IT=Image tube;} \\
\multicolumn{6}{l}{\hspace*{2.5em} CCD=Charge Coupled Device; Ret=Reticon} \\

\end{tabular}
\end{table}

\clearpage
\subsection{Figure Captions}

\begin{figure}[p]
\vspace{8in}
\includegraphics{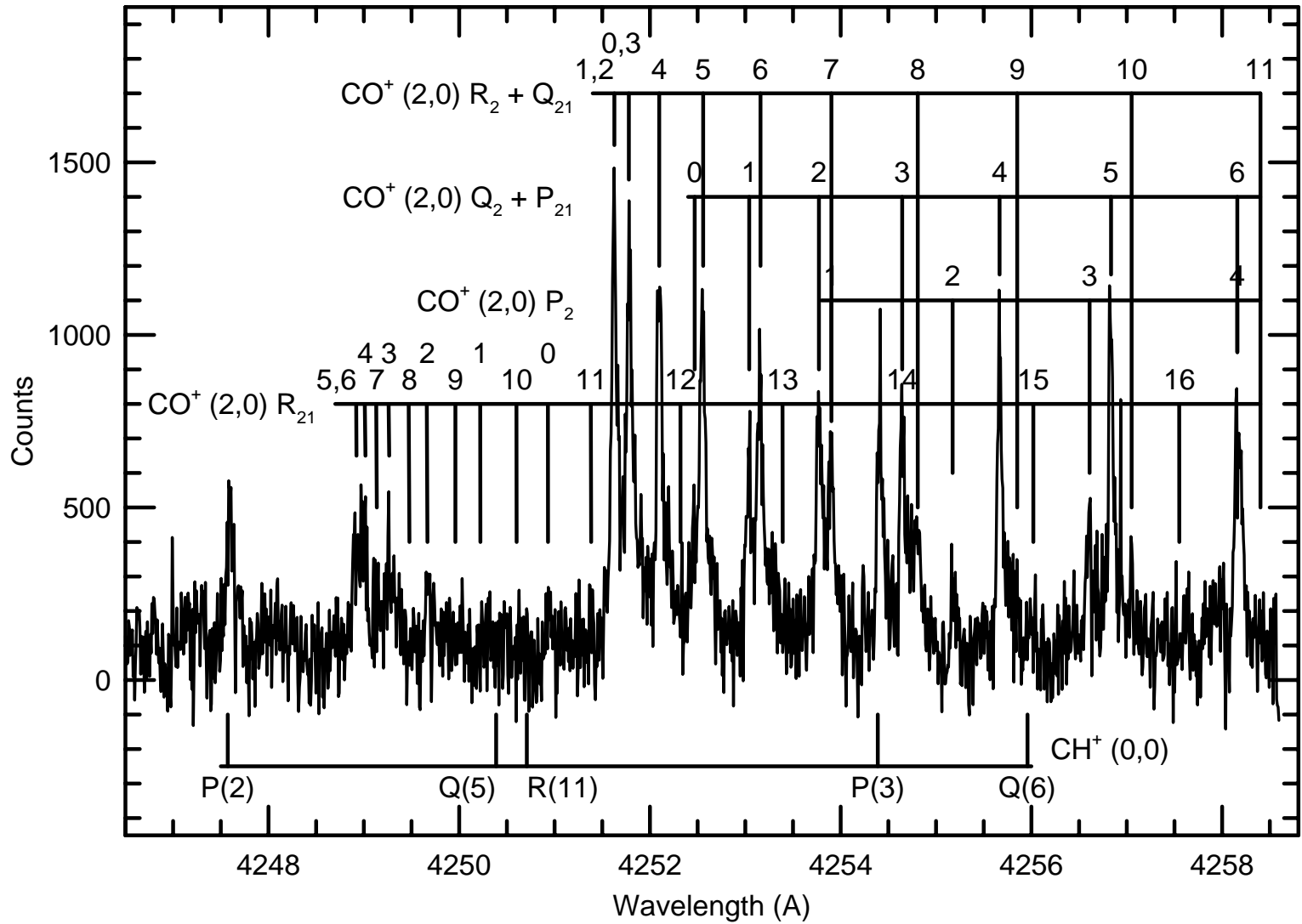}
\caption{Cochran, Cochran and Barker}\label{haleboppco+}
\end{figure}

\noindent
{\bfseries Figure~\ref{haleboppco+}:}
The spectral region of the CO$^+$ (2,0) and CH$^+$ (0,0)
bands for Hale-Bopp.  The positions of lines within this bandpass
are marked, though not all marked lines are present.  The pattern
of detected vs. non-detected lines can be explained by the excitation
levels of these molecules.

\begin{figure}[p]
\vspace{8in}
\includegraphics{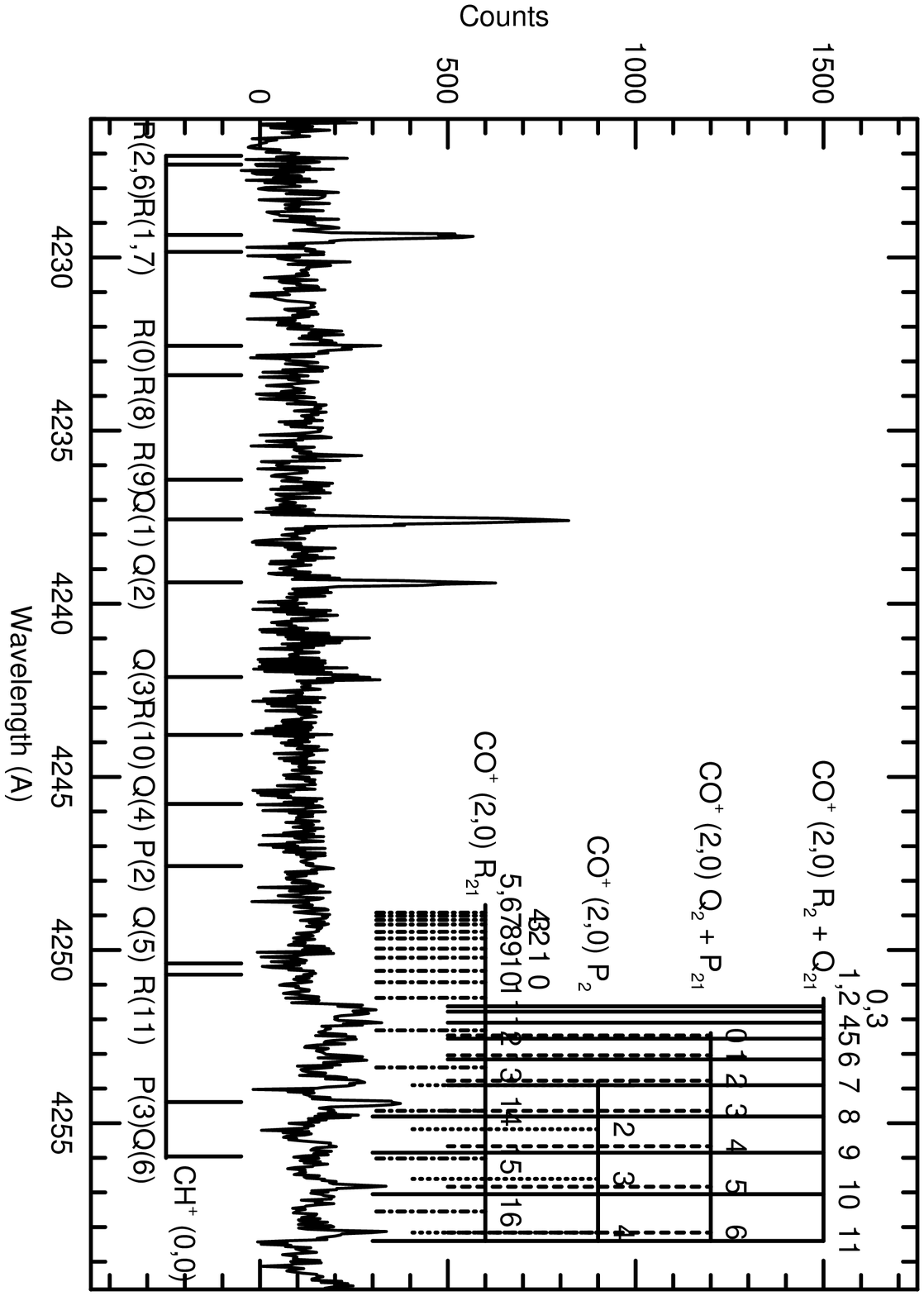}
\caption{Cochran, Cochran and Barker}\label{devicoco+}
\end{figure}

\noindent
{\bfseries Figure~\ref{devicoco+}:}
The spectral region of the CO$^+$ (2,0) and CH$^+$ (0,0)
bands for deVico.  The spectral orders are longer for the R=60,000
mode so more CH$^+$ lines were detected.  Note that CH$^+$ is
stronger relative to CO$^+$ in deVico than in Hale-Bopp.

\begin{figure}[p]
\vspace{8in}
\includegraphics{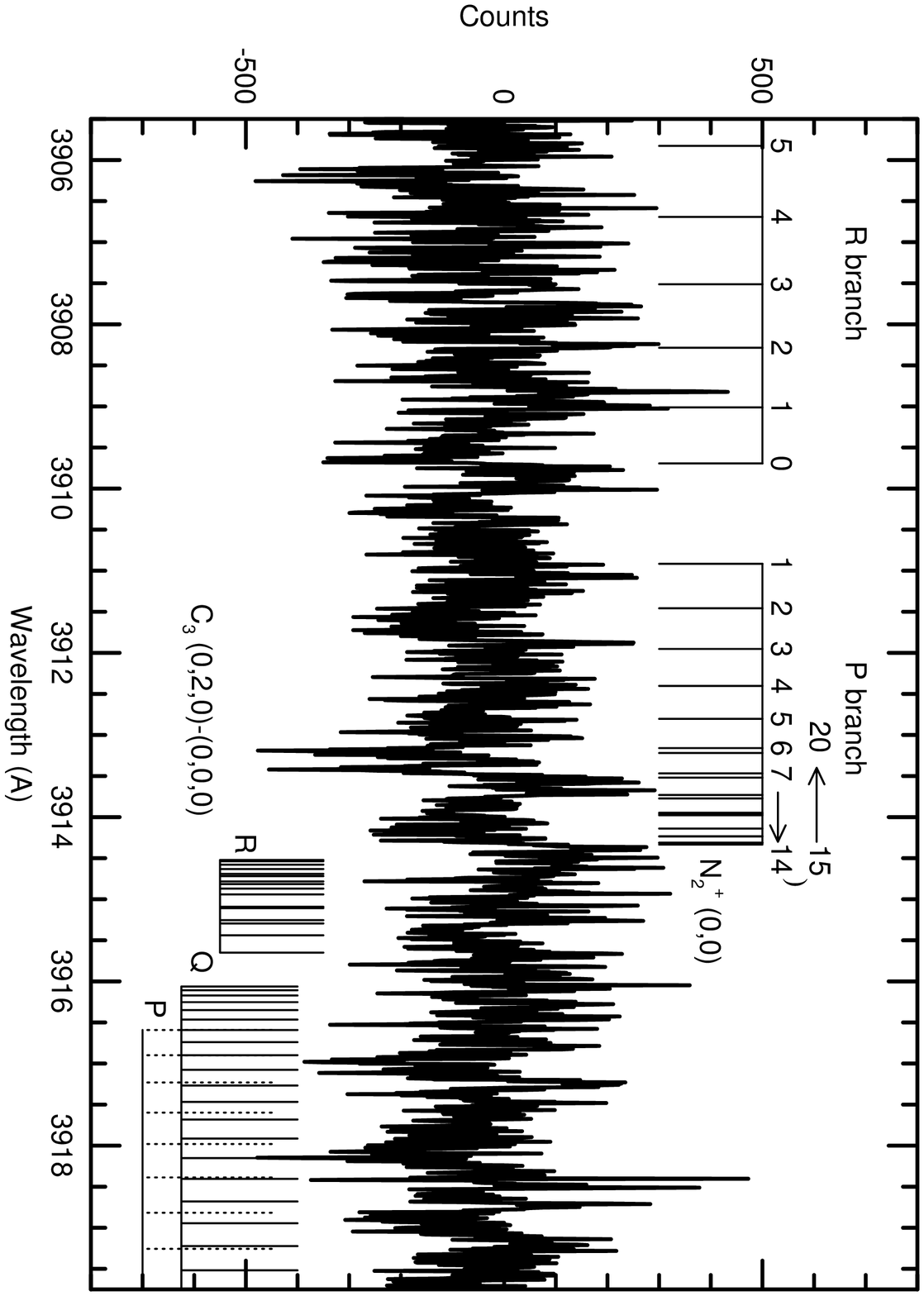}
\caption{Cochran, Cochran and Barker}\label{haleboppn2+}
\end{figure}

\noindent
{\bfseries Figure~\ref{haleboppn2+}:}
The spectral region of the N$_2^+$ (0,0) band for Hale-Bopp.
In addition, the C$_{3}$ (0,2,0)--(0,0,0) band is in this
spectral region.  The expected positions of lines for both of these
bands are marked.  There appears to be a feature at 3914.5\AA, which
we tentatively attribute to C$_{3}$, \underline{not} N$_2^+$.

\begin{figure}[p]
\vspace{8in}
\includegraphics{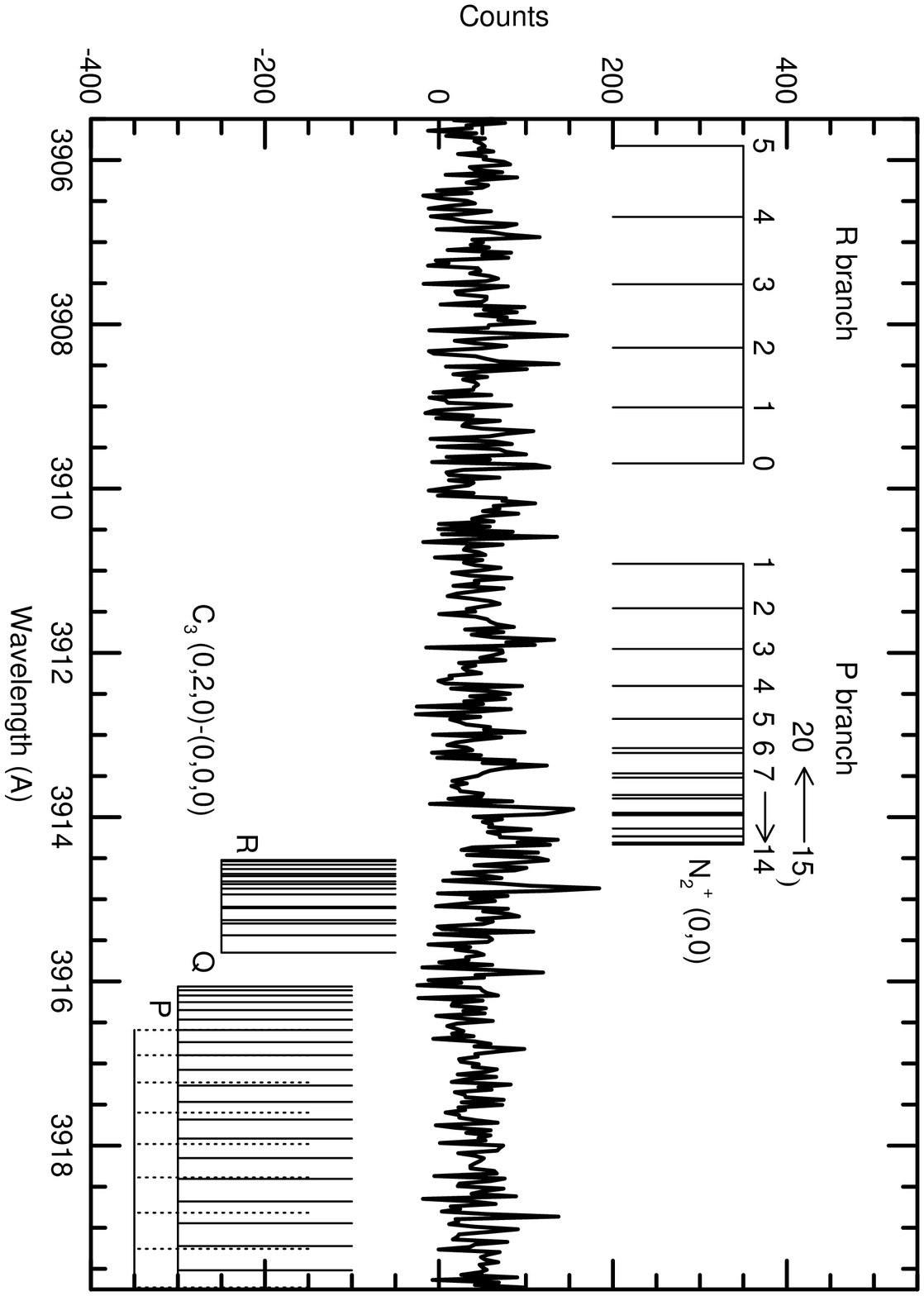}
\caption{Cochran, Cochran and Barker}\label{devicon2+}
\end{figure}

\noindent
{\bfseries Figure~\ref{devicon2+}:}
The spectral region of the N$_2^+$ (0,0) band for deVico.
As with Hale-Bopp, there appears to be a feature at 3914.5\AA.
However, we believe this feature is just noise.

\begin{figure}[p]
\vspace{8in}
\includegraphics{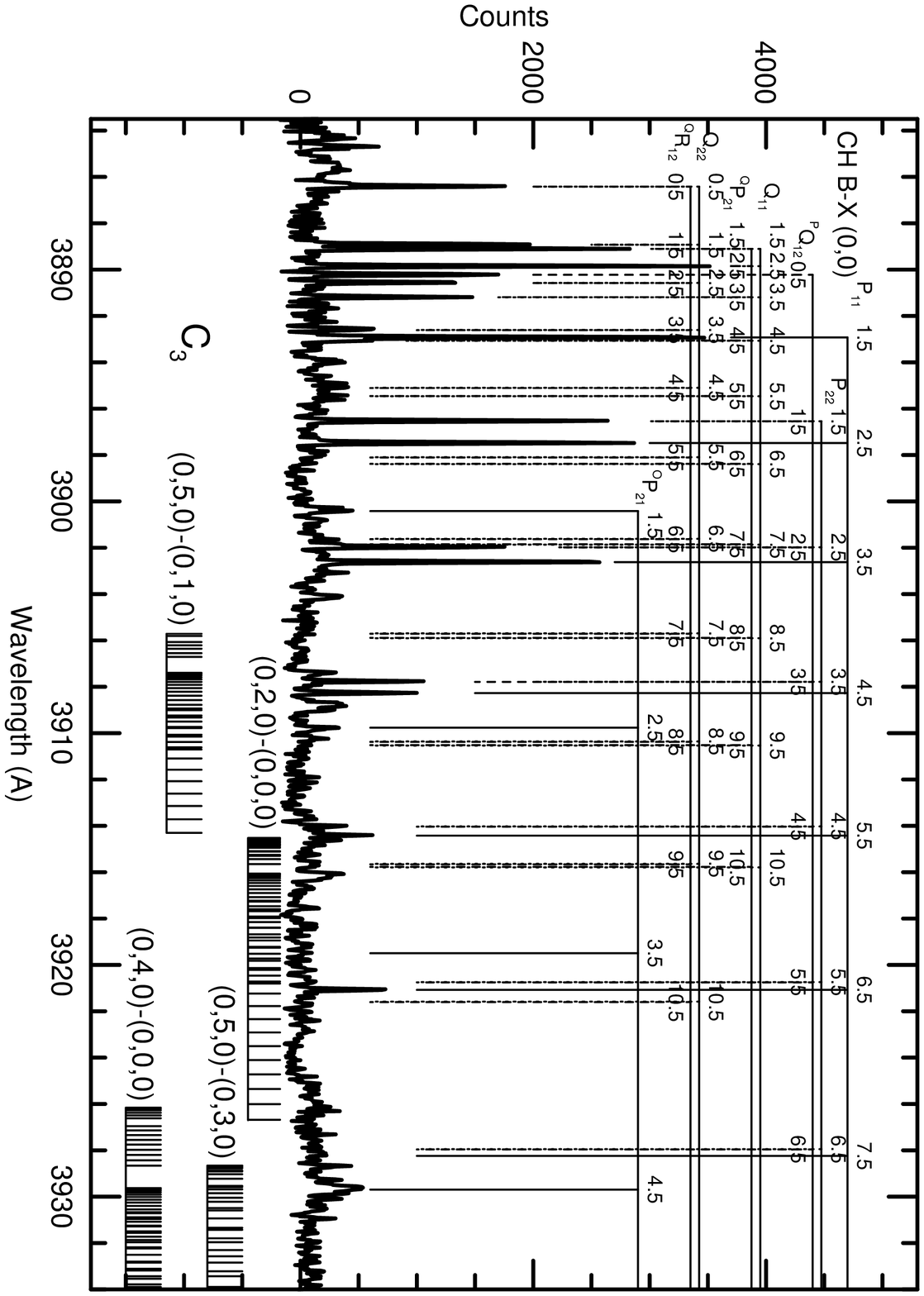}
\caption{Cochran, Cochran and Barker}\label{devicoopt}
\end{figure}

\noindent
{\bfseries Figure~\ref{devicoopt}:}
An optocenter spectrum of deVico, including the region
of the (0,0) band of N$_2^+$.  Several C$_{3}$ bands and the
CH B--X (0,0) band are identified.  Additional, unidentified, C$_{3}$
bands are probably present.  We show this spectrum to show the
abundance of C$_{3}$ in this comet.

\begin{figure}[p]
\vspace{8in}
\includegraphics{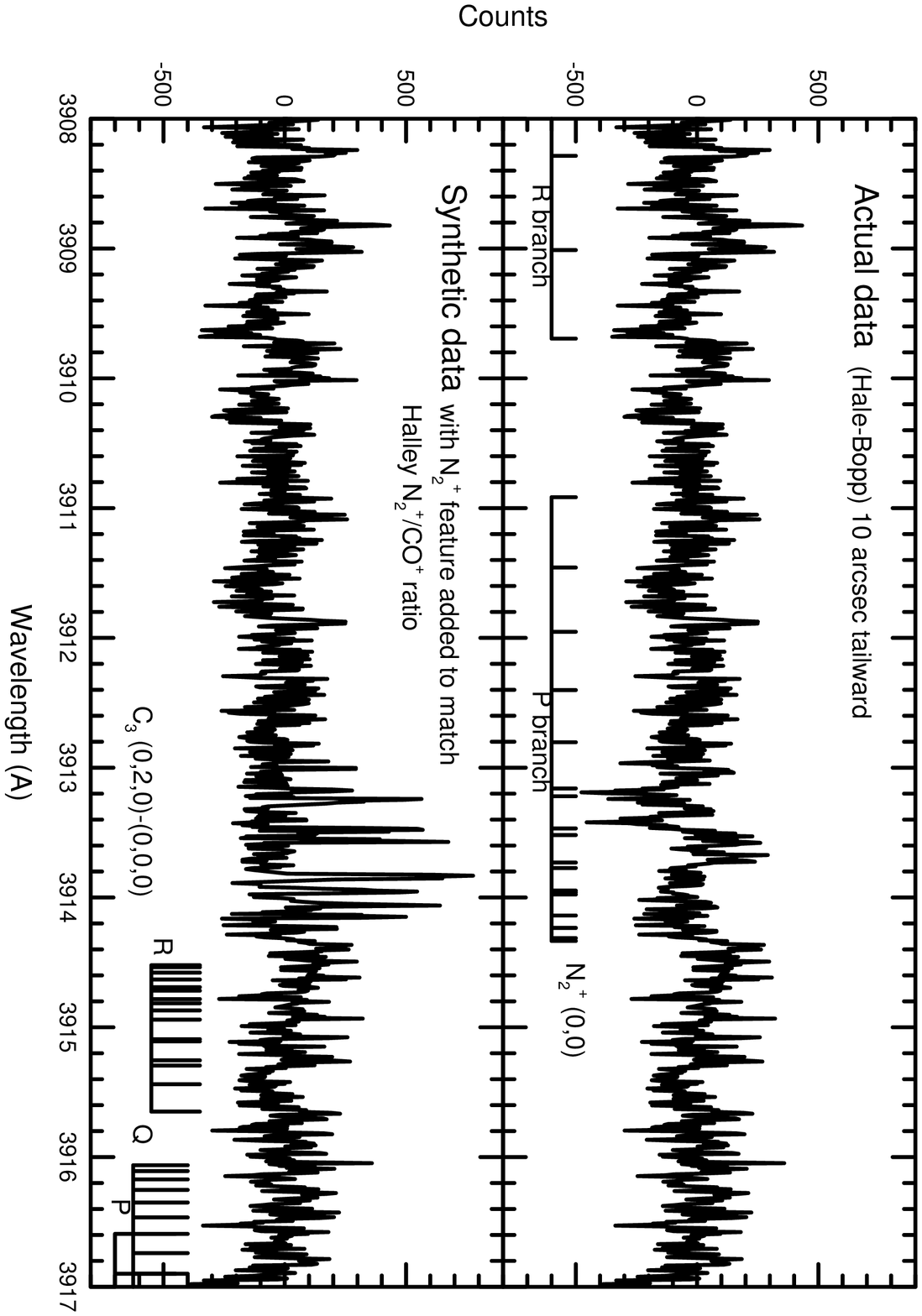}
\caption{Cochran, Cochran and Barker}\label{fake}
\end{figure}

\noindent
{\bfseries Figure~\ref{fake}:}
Simulated N$_2^+$ data.  The upper panel shows an
actual spectrum of Hale-Bopp.  The positions of the potential N$_2^+$ lines
are marked beneath it.  The lower panel shows this same spectrum 
but with a ``fake'' band replacing the data between 3913 and 3914\AA.
The fake band would have enough counts so that the ratio of
N$_2^+$/CO$^+$ would equal the value detected for Halley.  
In the real data, there are some upward excursions which do not correspond
to any N$_2^+$ lines.  Even integrating just these upward excursions
yields only 1/4 the necessary counts.

\begin{figure}[p]
\vspace{8in}
\includegraphics{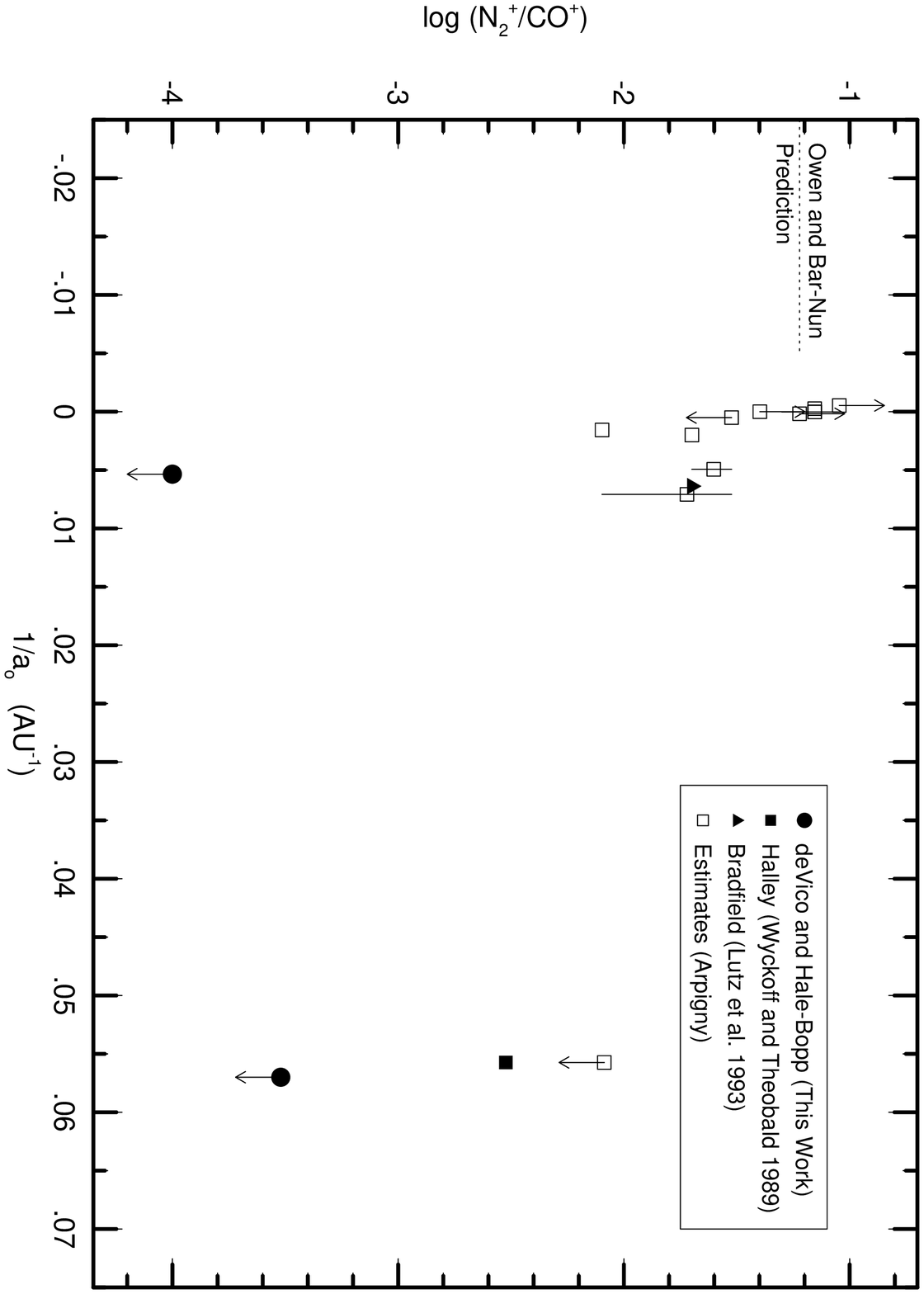}
\caption{Cochran, Cochran and Barker}\label{values}
\end{figure}

\noindent 
{\bfseries Figure~\ref{values}:}
Values for N$_2^+$/CO$^+$ as a function of 1/a$_o$.  The various
ratios and limits are plotted in order to examine whether a trend
exists with dynamical age of the comets.  The prediction for this
ratio of Owen and Bar-Nun (1985a) is shown as a dotted line.  
The open symbols are values derived from estimates of features in
photographic and digital spectra,
while the closed symbols represent measured digital spectra.  See text for
a discussion.

\clearpage

\end{document}